\documentclass{aa}
\usepackage{graphicx}
\usepackage{amssymb}
\def\omc{$\Omega/\Omega_{\rm{c}}$}

\begin{document}
%
\title{A study of the B and Be star population in the field of the LMC 
open cluster NGC2004 with VLT-FLAMES}
\titlerunning{B and Be stars in the field of the LMC-NGC2004}

\author{
C. Martayan \inst{1}
\and  A.M. Hubert  \inst{1}
\and  M. Floquet \inst{1}
\and  J. Fabregat \inst{2}
\and  Y. Fr\'emat \inst{3}
\and  C. Neiner \inst{4,1}
\and  P. Stee \inst{5}
\and  J. Zorec \inst{6}
}
\offprints {C. Martayan}
\mail{christophe.martayan@obspm.fr}
\institute{
GEPI, UMR8111 du CNRS, Observatoire de Paris-Meudon, 92195 Meudon Cedex, France
\and Universidad de Valenc\'ia, Instituto de Ciencia de los Materiales,
P.O. Box 22085, 46071 Valenc\'ia, Spain
\and Royal Observatory of Belgium, 3 Avenue Circulaire, 31180 Brussels, Belgium
\and Instituut voor Sterrenkunde, KU Leuven, Celestijnenlaan 220B,
3001 Belgium
\and Observatoire de la C\^ote d'Azur, avenue Nicolas Copernic, 06130
Grasse, France
\and Institut d'Astrophysique de Paris (IAP), 98bis boulevard Arago,
75014 Paris, France
}
\date{Received /Accepted}
\abstract{
Observations of hot stars belonging to the young cluster LMC-NGC2004  and its
surrounding region have been obtained with the VLT-GIRAFFE facilities in
MEDUSA mode. 25 Be stars were discovered; the proportion of Be stars compared
to B-type stars is found to be of the same order in the LMC and in the Galaxy
fields. 23 hot stars were discovered as  spectroscopic binaries (SB1 and SB2),
5 of these are found to be eclipsing systems from the MACHO database,  with
periods of a few days. About 75\% of the spectra in our sample are
polluted by hydrogen (H$\alpha$ and H$\gamma$), [\ion{S}{ii}] and
[\ion{N}{ii}] nebular lines. These lines are typical of \ion{H}{ii} regions. 
They
could be associated with patchy nebulosities with a bi-modal distribution in
radial velocity, with higher values (+335 km~s$^{-1}$) preferentially seen
inside the southern part of the known bubble LMC4 observed in \ion{H}{i} at 21
cm.
\keywords{Stars: early-type -- Stars: emission-line, Be -- Galaxies:
Magellanic Clouds -- Stars: binaries: spectroscopic -- Stars: binaries:
eclipsing -- ISM: nebular lines and bands}
}
\maketitle

\section{Introduction}
The origin of the Be phenomenon, i.e. periods of spectral emission due to the
presence of a circumstellar envelope around Be stars, is still debated. Fast
rotation seems to be  a major factor in triggering this phenomenon.  
To understand the Be phenomenon, it is important  to know at which phase of
stellar evolution on the main sequence
(MS) it appears. According to a statistical study of Be stars in clusters, 
Fabregat \& Torrej\'on (2000) concluded that it may occur in the
second half of the MS phase. Taking into account effects due to fast
rotation, Zorec (2004) concluded that the appearance of the Be
phenomenon among field early-type stars is probably mass-dependent and that it
may appear at any time during the MS phase. 
 However, according to Maeder \&
Meynet (2000), during the MS evolution rotational speed rises in
terms of critical velocity fraction (\omc), on the one hand, and stellar rotation is faster
as the metallicity of the protostellar clouds is lower, on the other hand.
Thus, the formation of rapid rotators and consequently of Be stars is favoured 
in regions of low metallicity.
Moreover, the increase of speed in \omc \ during the MS evolution 
also favours  the formation of rapid rotators, 
and consequently  of Be stars, at the later MS stages.

To investigate this issue and to determine the evolutionary state of Be stars,
(i) fundamental parameters of B and Be star populations located in
regions of different metallicity must be obtained, and (ii) the
specific properties of emission lines need to be determined. The new
instrumentation FLAMES-GIRAFFE installed at the VLT is particularly well suited
to obtain the medium resolution spectra of large samples of B and Be
stars, in low metallicity regions such as the Magellanic Clouds that 
are needed to achieve these goals. 

In this first paper we present an overview of spectroscopic results for 176
early-type stars observed in LMC NGC2004 and its surrounding field with
VLT-FLAMES. The observations and the reduction process are described in Sect.
2. In Sect. 3 we present the detection of new Be stars (Sect. 3.1), a
study of nebular lines (hydrogen, [\ion{N}{ii}] and [\ion{S}{ii}]) that
pollute the stellar spectra and apparently 
originate from filamentary gaseous structures (Sect. 3.2), and the discovery of spectroscopic
binaries (Sect. 3.3). Results are summarized in Sect. 4.

\section{Observations and reduction}
To obtain spectra of a significant sample of the B star population in
the young cluster LMC NGC2004 and its surrounding field, we made use
of the ESO VLT-FLAMES facilities, as part of the Guaranteed Time Observation
programs of the Paris Observatory (P.I.: F. Hammer). The multifibres
spectrograph VLT-FLAMES was used in MEDUSA mode (132 fibres) 
at medium resolution. 

The use of the setup LR02 (396.4--456.7nm, hereafter blue spectra) is
intended for determination of fundamental parameters, while the LR06 setup
(643.8--718.4 nm, hereafter red spectra) is used to identify Be stars 
and to study the H$\alpha$ line emission characteristics. Spectral
resolution is 6400 for LR02 and 8600 for LR06. Observations (ESO runs
72.D-0245B and 73.D-0133A) were carried out in 2003 on  November 24 (blue
spectra) and 28 (red spectra) and in 2004 on April 12 (blue spectra) and 14
(red spectra). The observational seeing ranged from 0.4 to
0.8$\arcsec$. 

Observations with the VLT-FLAMES facilities are performed with blind pointing
and therefore require very accurate astrometric data (better than
0.3$\arcsec$) for correct positioning of the thin fibres 
($\phi$ = 1.2$\arcsec$) on
the targets. To identify B-type stars in the selected field and to obtain their
position with the requested accuracy we used the EIS pre-FLAMES (LMC33) survey
images provided by ESO (see http://www.eso.org/science/eis/eis\_home.html). 
For all the sources extracted from EIS images, the astrometric and photometric
data were extracted with the SExtractor software (Bertin \& Arnouts
1996). 

\begin{figure}[]
	\centering
	\resizebox{\hsize}{!}{\includegraphics{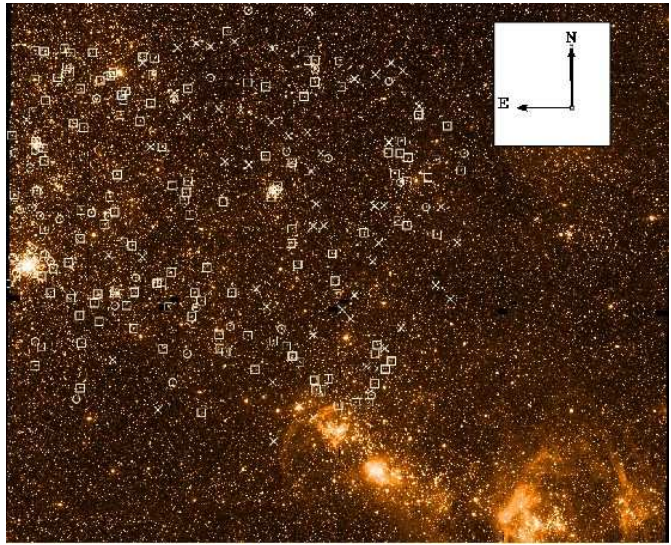}}
	\caption{The LMC33 field from EIS pre-FLAMES survey. Circles
	show Be stars in our sample. Squares show O, B and A-type
	stars in our sample and crosses indicate the sky fibre positions. The
	\ion{H}{ii}
	region LHA 120-N51A can be seen in the southwest corner of the field.}
	\label{figure0}
\end{figure}

The accuracy of B and V magnitude determination is $\sim$~0.2 mag.
B-type candidates were pre-selected with 14~$\le$~V~$\le$~18 and a colour
index B$-$V~$\le$~0.35, keeping in mind the intrinsic value E(B-V)=0.1 (Keller
1999) for the LMC. In addition, we chose to observe a list of known Be stars
from Keller et al. (1999). All our stellar and sky fibre targets are plotted
in Fig.~\ref{figure0}.

A sample of 177 stars was observed within the two observing
runs. 25 sky spectra were simultaneously obtained. 16 stars that were
found to have variable radial velocity (RV) in a 4-day interval in
November 2003 were observed again in April 2004. 

As the V magnitude of the selected targets ranges from 13.7 to 17.8 mag, we
chose a 2 and 1.5-hour integration time for the setups LR02
and LR06 respectively. This corresponds to a signal to noise ratio S/N
$\sim$120 in the blue spectral range and $\sim$80 in the red
spectral range for stars with V $\sim$ 16.4 mag.  Under these
conditions the S/N ratio of spectra obtained in the blue region varies
from $\sim$20 for the fainter objects to $\sim$150 for the
brighter ones.

Bias, flat-fields and wavelength calibration exposures (ThAr) were
obtained for each stellar exposure and used to reduce the spectra. The data
reduction was performed with the dedicated software GIRBLDRS developed at the
Geneva Observatory (see http://girbldrs.sourceforge.net) and several tasks of
the IRAF\footnote{IRAF is distributed by the National Optical Astronomy
Observatories, which is operated by the Association of Universities for
Research in Astronomy (AURA), Inc., under cooperative agreement with the
National Science Foundation.} package for extraction, calibration and sky
correction of the spectra were also used.

The observed field (25\arcmin~in diameter) is centered at $\alpha$(2000) = 05h
29min 00s and $\delta$(2000) = $-$67$^{\circ}$ 14$\arcmin$ 00$\arcsec$. 
Besides the young cluster NGC 2004, this field  contains several high-density 
groups of stars (KMHK943, 971, 963, 991, 988 and BSDL 2001). It also includes 
the southern part  of the LMC4 (SGS 11) \ion{H}{i} supergiant shell 
(Bica et al. 1999) 
and the northern extension of the \ion{H}{ii} region LHA 120-N 51A. 
The rim of LMC4 
is spotted with \ion{H}{ii} regions rich in young stars and 
OB associations 
(Domg\"{o}rgen et al. 1995).

\section{Results}

From the VLT-FLAMES observations with the setups LR02 and LR06 we
confirm the B spectral type for 168 of the 177 observed stars. Of the remaining
9 objects, 6 are O stars (presence of strong \ion{He}{ii} lines), 2 are A stars
and 1 is a cool star. 

After subtraction of the sky line contribution,  it
appeared that the spectra of more than two thirds of the 177
stars, as well as the totality of the sky spectra, are contaminated by nebular
lines. This contamination is particularly visible in the H$\alpha$
line. Depending on the intensity level of the nebular H$\alpha$ line, a weak
nebular contribution can also be detected in the H$\gamma$ and
H$\delta$ line profiles, as well as forbidden lines of [\ion{N}{ii}] at 6548
and 6583 \AA\ and [\ion{S}{ii}] at 6717 and 6731 \AA. When the nebular
H$\alpha$ line is very strong, a narrow emission component can also be observed
in the \ion{He}{i} line at 6678 \AA. Thus, in order to produce a robust
identification of stars with the Be phenomenon, we have to disentangle the
circumstellar (CS) line emission component from emission produced by nebular
line in polluted spectra.

\subsection{Be stars}

Up to now most Be stars in the LMC have been investigated with
photometric surveys. In particular, Keller et al. (1999, 2000) detected
numerous Be stars in NGC2004 with broad band photometry and a specific narrow
filter centered at H$\alpha$. In this paper, medium
resolution spectroscopy of Be stars in the field of NGC2004 in the LMC is
presented for the first time. 

Among the 25 known Be stars, called KWBBe in Keller et al. (1999), we confirm
the Be character of 22 stars. However, the stars KWBBe554, KWBBe993
and KWBBe1169 did not present emission in the H$\alpha$ line at 
the epochs
of our observations, which could be due to the transient nature of the Be
phenomenon.
The stars KWBBe203 and KWBBe287, showing a discrepancy between blue and red
radial velocities in November 2003, were again observed in April 2004.
Both stars display conspicuous variation in the intensity of H$\alpha$ over 5
months (see Table~\ref{table1}).

Moreover, among the B-type stars observed, 25 objects were found to be 
new Be stars (MHFBe in Table~\ref{table1}). This raises the total of
Be stars in our sample to 47. Of this sample of Be stars, the 
H$\alpha$ emission
line is single-peaked for 19 objects and double-peaked for 28 objects, 
while 12 of
the latter are shell stars (after correction of the nebular 
contribution in the H$\alpha$ line profile).
26 Be stars show \ion{Fe}{ii} in emission and 9 show red \ion{He}{i} lines at 6678 and
7065 \AA \ in double emission (see Table 1).

\begin{table*}[!]
	\caption{List of Be stars and H$\alpha$ emission line parameters.
KWBBe names from Keller et al. (1999) or our MHFBe catalogue number is given in
column 1. In the last column some complementary indications on the spectrum
are given: `cl' the star belongs to a cluster, `bin' the star is a binary;
`sh' the star is a shell star; `\ion{Fe}{ii}' and `\ion{He}{i}' indicate that
the corresponding line presents emission components, `w' means that the
corresponding emission lines are weak, `noneb' no nebular lines have been
detected; `$\alpha$(CS+neb)' the nebular line is present at H$\alpha$ but
cannot be disentangled from the CS emission and `$\alpha$CS+$\alpha$neb' the 
H$\alpha$ CS and nebular lines can be disentangled.}
\begin{tabular}{@{\ }ll@{\ }lllllll@{\ }l@{\ }}
\hline
\hline		
Star & $\alpha$ & $\delta$ & V & EW$\alpha$  & Imax & I(V) & I(R) & FWHM  &
Remarks \\
     & (2000)& (2000)&   &{\AA}&      &      &      & (km~s$^{-1}$)&   \\
\hline
   	KWBBe44    & 05 30 45.049 & -67 14 26.13 & 13.7 & 30.94 & 7.31
&  &  &180& $\alpha$(CS+neb), \ion{Fe}{ii}\\
	KWBBe75    & 05 30 37.690 & -67 17 39.50 & 14.4 & 65.48 & 12.97 &
&  & 246 & cl, $\alpha$(CS+neb), \ion{He}{i}, \ion{Fe}{ii}\\
	KWBBe91    & 05 30 44.284 & -67 17 22.71 & 14.4 & 34.64 & 5.94 &
&  & 284 & cl, $\alpha$(CS+neb), \ion{Fe}{ii} \\
	KWBBe152   & 05 30 24.391 & -67 14 55.49 & 15.6 & 68.92 & 10.16 &
&  & 331 & noneb, \ion{Fe}{ii} \\
	KWBBe171   & 05 30 36.340 & -67 16 51.00 & 15.5 & 16.31 &  & 2.92
&2.90 & 350 & cl, noneb, \ion{He}{i} \\
	KWBBe177   & 05 30 39.582 & -67 16 49.52 & 15.3 & 21.84 &  & 2.75 &
2.41 & 492 & cl, sh?, $\alpha$CS+$\alpha$neb, \ion{He}{i}, \ion{Fe}{ii}w \\
	KWBBe203   & 05 30 48.703 & -67 16 49.31 & 15.3 & 41.96 & 7.76 &
&  & 264 & cl, $\alpha$(CS+neb), \ion{Fe}{ii} \\
	            &              &              & 15.3 & 47.79 & 8.61 &
&  &  261 &  \\
    KWBBe276   & 05 29 59.358 & -67 14 46.48 & 15.8 & 51.45 &  &
6.06 & 6.36 & 406 & sh, $\alpha$(CS+neb), \ion{Fe}{ii} \\
	KWBBe287   & 05 30 08.074 & -67 14 36.25 & 15.8 & 42.58 &  & 7.05 &
4.93 &  & sh, noneb, bin, \ion{Fe}{ii} \\
	           &              &              & 15.8 & 36.06 &  & 6.07 &
4.92 &  &  \\
    KWBBe323   & 05 30 31.976 & -67 16 40.26 & 16.2 & 49.69 & 8.55
&  &  & 264 & cl, $\alpha$(CS+neb), \ion{Fe}{ii} \\
	KWBBe342   & 05 30 38.630 & -67 16 23.00 & 16.0 & 19.14 &  & 2.80 &
2.91 & 394 & cl, noneb \\
	KWBBe344   & 05 30 38.859 & -67 14 21.13 & 15.9 & 34.53 & 6.89 &
&  & 240 & noneb, \ion{Fe}{ii} \\
	KWBBe347   & 05 30 39.900 & -67 12 21.01 & 16.0 & 15.96 &  & 2.38 &
2.37 & 451 & noneb \\
	KWBBe374   & 05 30 47.799 & -67 11 36.43 & 16.4 & 33.42 & $\sim
4.15$ &  &  & 435 & $\alpha$CS+$\alpha$neb \\
	KWBBe579   & 05 30 27.610 & -67 13 0.209 & 16.5 & 39.51 &  & 4.83 &
4.74 & 445 & cl, sh, $\alpha$(CS+neb), \ion{Fe}{ii} \\
	KWBBe622   & 05 30 36.223 & -67 13 27.70 & 16.9 & 21.48 &  & 4.49 &
4.77 & 334 & $\alpha$(CS+neb)? \\
	KWBBe624   & 05 30 37.040 & -67 21 02.62 & 16.7 & 43.33 &  & 6.13 &
5.58 & 407 & sh, noneb, \ion{Fe}{ii}w \\
	KWBBe874   & 05 29 55.846 & -67 19 14.26 & 17.5 & 55.67 & 9.68 &
&  & 284 & $\alpha$(CS+neb), \ion{Fe}{ii}w \\
	KWBBe1055  & 05 30 29.265 & -67 17 01.65 & 17.4 & 54.76 & 6.37 &
&  & 398 & cl, sh?, $\alpha$(CS+neb), \ion{Fe}{ii} \\
	KWBBe1108  & 05 30 37.238 & -67 11 29.31 & 17.6 & 14.55 &  & 2.40 &
2.78 & 359 & cl, $\alpha$(CS+neb) \\
	KWBBe1175  & 05 30 46.960 & -67 17 39.30 & 17.7 & 40.05 &  & 4.94
&4.94 & 454 & cl, sh, noneb \\
	KWBBe1196  & 05 30 50.893 & -67 18 08.51 & 17.6 & 25.81 &  & 4.10 &
3.81 & 407 & cl, sh, noneb \\
	MHFBe55075      & 05 30 15.660 & -67 23 54.60 & 16.0 & 3.79  &  &
1.35 & 1.35 & 457 & $\alpha$CS+$\alpha$neb \\
	MHFBe55920      & 05 27 42.256 & -67 23 36.58 & 16.1 & 20.61 & 6.03
&  &  & 145 & $\alpha$(CS+neb), \ion{Fe}{ii} \\
	MHFBe59721      & 05 29 25.670 & -67 23 02.40 & 16.5 & 4.47  &  &
1.36 & 1.52 & 369 & $\alpha$CS+$\alpha$neb\\
	MHFBe66252      & 05 29 18.200 & -67 21 37.10 & 16.3 & 15.45 & 2.27
&  &  & 532 & $\alpha$(CS+neb), \ion{Fe}{ii} \\
	MHFBe72704      & 05 28 29.430 & -67 20 20.10 & 14.9 & 1.01  &  &
1.02 & 1.04 & 230 & noneb \\
	MHFBe73013      & 05 28 55.994 & -67 20 14.55 & 16.1 & 25.68 & $\sim
3.5$ &  &  & 397 & $\alpha$CS+$\alpha$neb, \ion{Fe}{ii}w \\
	MHFBe77796      & 05 29 13.609 & -67 19 18.70 & 14.9 & 16.84 &  &
3.85 & 3.85 & 258 & noneb, \ion{He}{i} \\
    MHFBe85028      & 05 29 45.000 & -67 17 50.50 & 16.1 & 6.50  &  &
1.63 &1.64 & 476 & sh, noneb, \ion{He}{i} \\
	MHFBe101350     & 05 30 34.640 & -67 14 45.30 & 15.3 & 10.81 &  &
2.52 & 2.57 & 315 & sh?,  noneb \\
	MHFBe103914     & 05 27 15.216 & -67 14 10.15 & 15.0 & 45.95 & 8.15
&  &  & 271 & $\alpha$(CS+neb), \ion{Fe}{ii} \\
	MHFBe107771     & 05 28 34.651 & -67 13 23.97 & 16.4 & 4.33  &  &
1.29 & 1.31 & 510  & cl, $\alpha$CS+$\alpha$neb \\
	MHFBe107877     & 05 28 34.260 & -67 13 25.90 & 15.2 & 36.77 &  &
4.68 & 5.10 & 373 & cl, noneb, \ion{He}{i}, \ion{Fe}{ii}  \\
	MHFBe108272     & 05 28 32.127 & -67 13 23.68 & 16.4 & 4.49  &  &
1.39 & 1.36 & 341 & cl, $\alpha$CS+$\alpha$neb \\
	MHFBe110827     & 05 27 28.582 & -67 12 56.81 & 14.9 & 34.21 &  &
3.51 & 	3.56 & 508 & cl, sh, noneb, \ion{Fe}{ii} \\
	MHFBe116297     & 05 28 06.021 & -67 11 47.43 & 16.1 & 10.56 &  &
2.21 & 2.17 & 331 & noneb \\
	MHFBe118313     & 05 26 55.945 & -67 11 27.09 & 15.0 & 59.50 & 10.15
&  &  & 265 & $\alpha$(CS+neb), \ion{He}{i}, \ion{Fe}{ii}w \\
	MHFBe118784     & 05 28 27.565 & -67 11 28.50 & 14.8 & 29.42 & 4.64
&  &  & 331 & noneb, \ion{He}{i}, \ion{Fe}{ii} \\
	MHFBe119521     & 05 30 35.252 & -67 11 19.10 & 16.3 & 9.67  &  &
1.73 & 1.74 & 574 & cl, noneb \\
    MHFBe132079     & 05 30 05.496 & -67 08 53.45 & 16.2 & 40.14 &
7.27 &  &  & 262 & noneb, \ion{Fe}{ii} \\
	MHFBe132205     & 05 29 25.872 & -67 08 53.93 & 14.7 & 32.90 & 7.14
&  &  & 212 & cl, $\alpha$(CS+neb), \ion{Fe}{ii} \\
	MHFBe136844     & 05 27 35.241 & -67 07 54.43 & 15.1 & 57.53 & 9.99
&  &  & 274 & $\alpha$(CS+neb), \ion{Fe}{ii} \\
	MHFBe137325     & 05 29 07.009 & -67 07 49.28 & 16.3 & 7.39  &  &
2.02 & 2.02 & 298 & noneb \\
	MHFBe138610     & 05 30 00.332 & -67 07 44.02 & 14.8 & 5.04  &  &
1.64 & 1.65 & 319 & cl, noneb \\
	MHFBe140012     & 05 28 13.281 & -67 07 20.77 & 14.9 & 14.59 &  &
2.55 & 2.55 & 429 & cl, sh?, noneb, \ion{He}{i} \\
	MHFBe155603     & 05 28 47.760 & -67 04 25.70 & 14.7 & 41.71 & 6.33
&  &  & 300 & noneb, \ion{Fe}{ii} \\
\hline
	\end{tabular}
	\label{table1}
\end{table*}

\subsubsection{Characteristics of the H$\alpha$ line emission}

\begin{figure}[tbp]
	\centering
	\resizebox{\hsize}{!}{\includegraphics{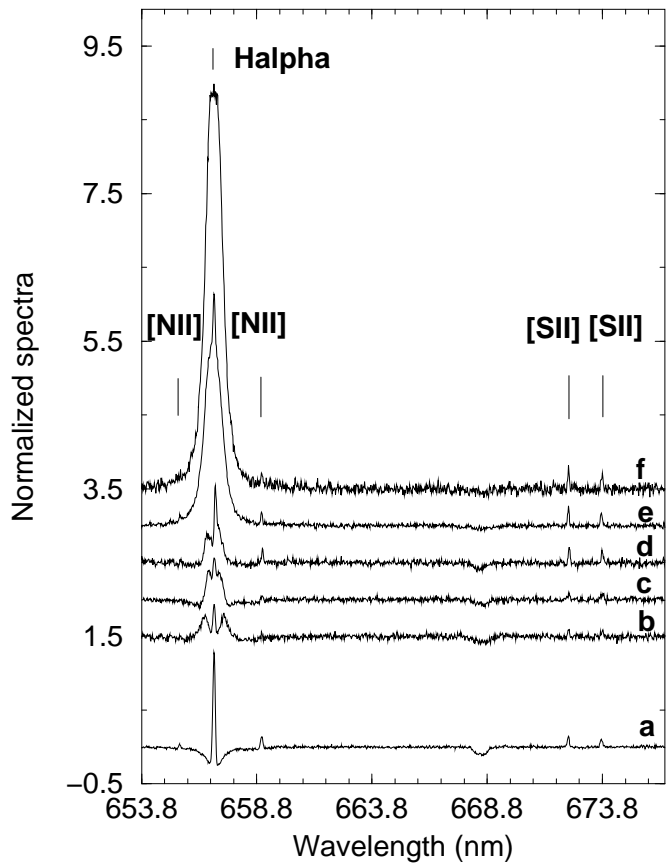}}
	\caption{Example of H$\alpha$ line spectral region showing nebular
emission lines of H$\alpha$, [\ion{N}{ii}] and [\ion{S}{ii}]. An example of a B
star spectrum (MHF54565) is labelled `a' and Be stars spectra
(MHFBe107771, MHFBe108272, MHFBe59721, MHFBe73013 and KWBBe1055) are
labelled `b', `c', `d', `e'  and `f', respectively.}
	\label{figure1}
\end{figure}

The equivalent width (EW$\alpha$), maximum intensity (Imax for a single peak,
and I(V) and I(R) for the violet and red peaks respectively in a
double peak emission) and the FWHM of the circumstellar H$\alpha$ emission
measured for each Be star are given in Table~\ref{table1}. The FWHM of the CS
H$\alpha$ emission line ranges between 145 and 575 km~s$^{-1}$.
The telluric lines like the nebular lines are resolved and the FWHM
of the H$\alpha$ nebular line is $\sim$55 km~s$^{-1}$ greater than the instrumental FWHM ($\sim$35 km~s$^{-1}$). 
An example of H$\alpha$ nebular line is shown in Fig.~\ref{figure1} with the telluric lines removed.
Thus, a Be star is easily distinguishable from a B star showing H$\alpha$ nebular contribution, except when the $V\!\sin i$ is
low and/or the CS emission is very weak. 

However, a fraction of Be stars shows a nebular contribution in their 
spectrum easily
detected by the presence of [\ion{N}{ii}] and [\ion{S}{ii}] lines, but not always
distinguishable in H$\alpha$ when the CS emission in this line is strong and
when the RVs of nebular and CS components are similar. This problem is chiefly
met in low or moderate $V\!\sin i$ Be stars. In these cases the nebular H$\alpha$
line emission contributes to overestimation of Imax and 
underestimation of the FWHM of the
CS emission line. To a lesser degree it also leads to overestimation of EW$\alpha$.

Examples of H$\alpha$ lines polluted by the nebular contribution are given in
Fig.~\ref{figure1}: \\
(a) shows an example of a B star (MHF54565) with strong nebular
emission lines;\\  
(b), (c), (d) illustrate Be stars (MHFBe107771, MHFBe108272 and
MHFBe59721) displaying 2 distinct V and R peaks plus a central nebular line;\\
(e) shows a Be spectrum (MHFBe73013) with nebular and CS emissions
severely blended into a single peak profile;\\
(f) illustrates the undetectable presence of a nebular line in the strong CS
H$\alpha$ emission of a Be star (KWBBe1055).

With the fibres located on sky positions (without stars in the
background),  we determine all parameters (FWHM, intensity) of the H$\alpha$
and [\ion{S}{ii}] nebular lines, and we obtain their intensity ratio:
[\ion{S}{ii}]/H$\alpha$~=~0.2 $\pm$ 0.1(see Section 3.2). 
As the intensity and RV of [\ion{S}{ii}] 6717 line could be measured in each 
Be spectrum,
and assuming that the nebular [\ion{S}{ii}]/H$\alpha$ ratio is the same for 
sky spectra and stellar spectra, 
it is possible to subtract the corresponding H$\alpha$ nebular contribution 
from the emission
line profile. Consequently, we can state precisely whether the H$\alpha$ CS emission 
is single, double or affected by a shell
component for KWBBe91, KWBBe177, KWBBe203, KWBBe374, KWBBe874, KWBBe1055, 
MHFBe73013,
MHFBe77796, MHFBe107771, MHFBe108272, MHFBe136844 and MHFBe155603.

For all Be stars presented in Table~\ref{table1}, we find no clear correlation 
between the FWHM and the EW of H$\alpha$ emission lines. Nevertheless, some
general trends can be derived:
\begin{itemize}
\item A single-peak emission is generally associated with large
EW$\alpha$ ($\geq$~20~\AA) and to narrow to moderate FWHM
($\lesssim$~350~km~s$^{-1}$).
\item  A double-peak emission without shell contribution is associated with a
small EW$\alpha$.
\item  A double-peak emission with a strong shell feature is associated with
a large FWHM ($\geq$~400~km~s$^{-1}$), whatever the EW$\alpha$ is.
\item \ion{Fe}{ii} emission lines are generally present in Be stars with 
EW$\alpha$~$>$~20~\AA.
\end{itemize}
These trends are agree fairly well with the conclusions derived by Dachs et al.
(1986) in their study of bright galactic Be stars.  Note, however, that
the minimum H$\alpha$ EW required for \ion{Fe}{ii} emission lines to become
visible in galactic Be stars is EW$\alpha$~$>$~7~\AA \ according to Hanuschik
(1987). Although their study and ours do not use the same \ion{Fe}{ii} lines,
the difference in the EW$\alpha$ limits between galactic and Magellanic Be stars
is consistent with their metallicity difference.

In addition, all the KWBBe stars show a strong H$\alpha$ emission (EW$\alpha$
$\geq 15$ {\AA}) compared with the MHFBe stars discovered in this paper. This is
probably a selection effect due to the photometric H$\alpha$ index criteria
used by Keller et al. (1999), which could not detect Be stars with weak
emission. 

The determination of fundamental parameters (temperature T$_{\rm eff}$, gravity
$\log$$g$, $V\!\sin i$) will be the subject of a forthcoming paper (Martayan et
al., in preparation). In this study however, we make use of our results on
$V\!\sin i$ values obtained by comparison between the observed spectra and a
grid of synthetic spectra in the spectral domain 4000-4500 \AA, which contains
5 strong helium lines. The adopted metallicity and abundances  for the
synthetic spectra are those derived for the LMC (Korn et al. 2002).

\begin{figure}[tbp]
	\centering
	\resizebox{\hsize}{!}{\includegraphics{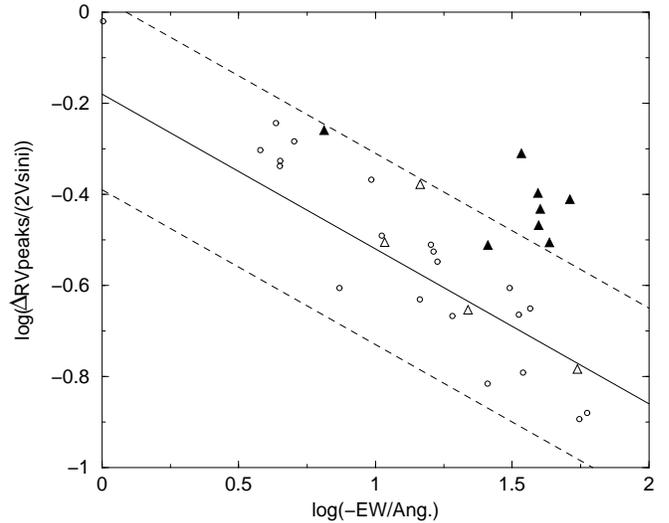}}
	\caption{Peak separation velocity of the H$\alpha$ CS emission line as
a function of its equivalent width. Filled triangles correspond to strong shell
Be stars, open triangles to weak shell Be stars and circles to ``normal" Be
stars. The solid line represents the mean correlation from Zamanov et al.
(2001), with dashed lines showing the dispersion.}
	\label{Beemission}
\end{figure}

For the H$\alpha$ line, we find a correlation on the one hand between $V\!\sin i$ and FWHM on
one hand, and on the other hand between ($\Delta RV{\rm peaks}$)/($2V\!\sin i$) (where $\Delta
RV{\rm peaks}$ is the peak separation velocity) and EW$\alpha$, as found for
galactic Be stars by Dachs et al. (1986) and Hanuschik et al. (1988), and
refined by Zamanov et al. (2001). This provides evidence for
a rotationally supported CS envelope. Taking into account the fact that the dispersion
of points for galactic  Be stars is similar to that for
LMC Be stars (see Fig. 2a, Zamanov et al 2001), the average slope of the peak
separation versus the equivalent widths of  galactic Be stars determined by
Zamanov et al. (2001) is comparable to ours (see
Fig.~\ref{Beemission}).

Nevertheless, we note that Be stars in the LMC with strong "shell"
(filled triangles in Fig.~\ref{Beemission}) are located on the right
in the diagram, above the mean correlation.  The increase of the peak
separation in those objects is an opacity effect (see eq.(7)
in Hanuschik et al. 1988). 
It indicates that Be stars in shell phases have denser disks.  This result
obtained for LMC Be stars confirms the similar result of Moujtahid
et al. (1999) in their study of photometric data translated into a
spectrophotometric  frame of galactic Be stars (see also Kogure 1989).

\begin{figure}[]
	\centering
	\resizebox{\hsize}{!}{\includegraphics{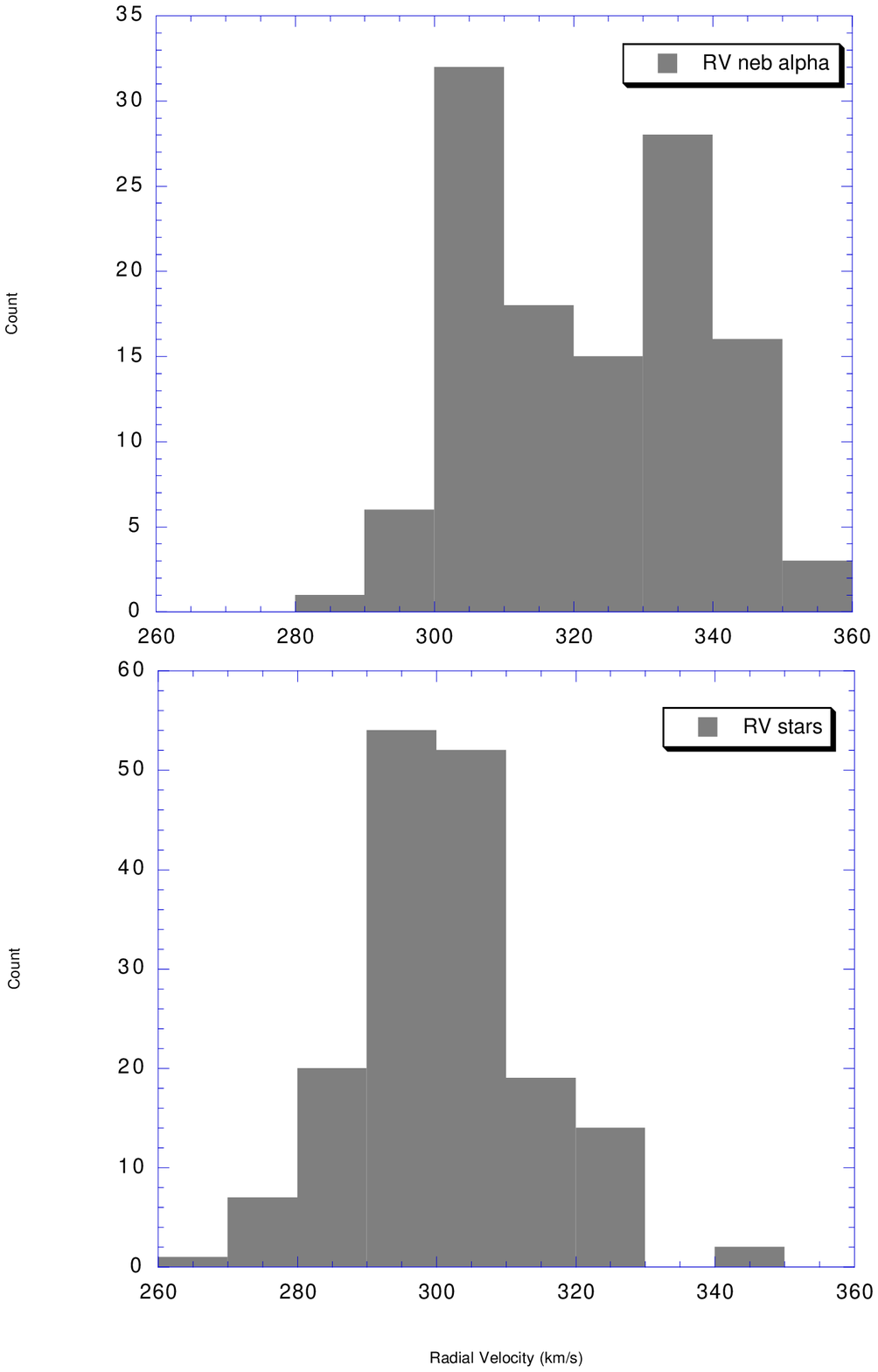}}
	\caption{Distribution of radial velocities for nebular lines (top)
	and observed stars (bottom).}
	\label{figure2}
\end{figure}
\begin{figure}[tbp]
	\centering
	\resizebox{\hsize}{!}{\includegraphics[angle=270]{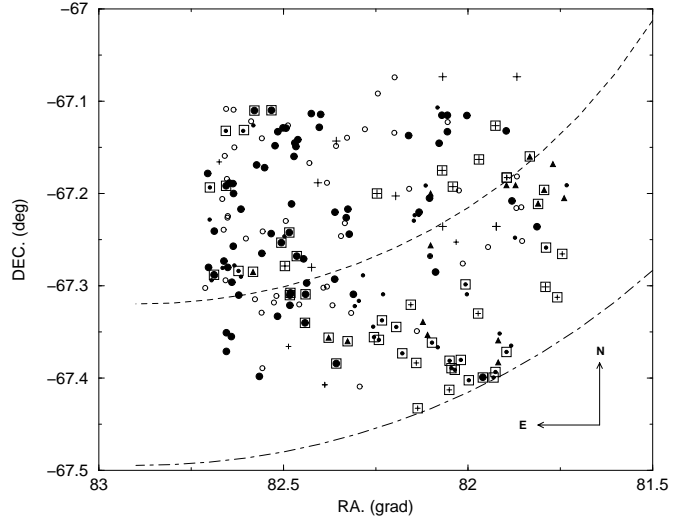}}
	\caption{Observed field of NGC2004 and its surrounding region showing
stars with no nebular line (small open circle), with nebular line with low RV
(small filled circle), intermediate RV (filled triangle) and high RV (large
filled circle). Crosses represent sky fibres showing nebular lines with low RV
(small) and high RV (large). Large open squares indicate that the intensity of
the H$\alpha$ nebular line is strong. The dashed and dotted-dashed lines
represent the southern limit of the LMC4 supergiant shell according to Kim et
al. (1999) and Meaburn (1980) respectively.}
	\label{figure3}
\end{figure}

\subsubsection{Proportion of Be stars}

We determine the percentage of Be stars found on the one hand in the observed
clusters and OB concentrations and on the other hand in the field close to
NGC2004. ``Be stars'' are all non-supergiant late
O, B and early A-type stars that have shown some emission in
their Balmer lines  at least once (Jaschek et al. 1981). This definition is also used here.

Excluding the area covered by the study of Keller et al. (1999), where we
observed 22 known KWBBe stars and 1 new Be star, we discovered 24
other new Be stars among the remaining 131 observed O, B, A-type stars. They
represent $\sim$18.3\% of this unbiased sample. 6 of these Be stars
seem to be located in OB concentrations or in small clusters, and 18 are
located in the field outside the clusters.

The percentages of Be stars in the clusters and in the field
surrounding NGC 2004, but excluding NGC 2004 itself and its stars, are thus the
following:
\begin{itemize}
\item 
In clusters and OB associations we find 7 Be stars and 12 normal
absorption-line O, B stars. We obtain a mean percentage N(Be)/N(B+Be)= 42\%
$\pm$ 28\%. The large error bar is due to the low number of observed stars in
clusters and associations.
\\
\item
For statistical analysis in the field of NGC2004, we use two
methods. We exclude the area covered by the study of Keller
et al. (1999). This represents 17 Be stars and 95 B stars, i.e. $\sim$15\% of
Be stars.  In the second method, we exclude only the clusters instead of the
entire area. This represents 28 Be
stars and 112 B stars, i.e. $\sim$20\% of the Be stars.
The mean percentage for the field of NGC2004:
N(Be)/N(B+Be)=17.5\% $\pm$ 2.5\%. \\

\end{itemize}

Due to the characteristics of the instrumentation, which do not make 
it possible to
position many fibres in a small angular size region ($\le 1 \arcmin$), 
and due to the
characteristics of our survey conducted in two single epochs and the variable
nature of the Be phenomenon, the above figures should be considered as a rough
estimate of the frequency of Be stars inside the clusters. 
Nevertheless, we can draw the following conclusions:

\begin{itemize}
\item The study by Fabregat (2003) shows a clear difference between the
Be star content of young galactic clusters compared with the content of the
surrounding galactic field. This implies that the Be star population in
co-eval stellar samples is not representative of the population in mixed-age
samples in the field. Our results for Be star contents in clusters and in the
field of the LMC are similar to those of previous studies in the Galaxy.
Moreover, the percentage of Be stars we determined is close to the
estimate made by Keller et al. (2000) in young LMC clusters.
\\
\item The frequency of Be stars in the LMC field is comparable to 

the mean frequency in the galactic field, which is about 20\% for the early B
subtypes (Zorec \& Briot, 1997). This result is in agreement with 
that of
Keller et al. (1999), who found that the fraction of Be stars compared
with B-type stars lies between 10 to 27\% in the field population around the LMC
clusters they studied and is 11\% for the field around the cluster NGC2004. As
the metallicity of the LMC is significantly lower than the galactic 
metallicity, this
result casts doubt on the dependence of Be frequency on metallicity as
proposed by Maeder et al. (1999).
\end{itemize}
\subsection{Nebular lines}

\begin{table*}[!]
	\centering
	\caption{Spectroscopic binary stars in the NGC2004 field. RVs are
expressed in km~s$^{-1}$ and have an accuracy of $\pm10$ km~s$^{-1}$. 
In column 9, some indication of variability
in MACHO data is given: 'no var' if the star shows no photometric variability greater than 0.05 mag., 
'no data' if the star is not found in the database; 
the value of the orbital period we determined for 5 systems is expressed in days with 
an accuracy of 0.001 d. Remarks are given in the last column: 
'SB2' for spectroscopic binaries with 2 components and 'SB?' for a potential
spectroscopic binary (the other binaries in the Table are SB1). 
}
	\begin{tabular}{llllllllll}
\hline
\hline
Binary & $\alpha$ & $\delta$ & V &
RV\small{2003}&RV\small{2003}&RV\small{2004}&RV\small{2004}& MACHO &
Remarks  \\
     &\small{(2000)}&\small{(2000)} & &\small{11/24}& \small{11/28}
&\small{04/12} & \small{04/14}& & \\
\hline
MHF64847 &05 27 32.04 & -67 21 53.20 & 15.13 & +396 & +264 & +364 & +221 & no var &  \\
MHF65587 &05 28 23.46 & -67 21 41.30 & 16.01 & +322 & +276 & +297 & +273 & no var &  \\
MHF71137 &05 28 46.65 & -67 20 40.90 & 15.62 & +308 & +265 & +263 & +279 & no var &  \\
MHF79301 &05 29 11.05 & -67 18 58.90 & 15.61 & +316 & +296 & +313 & +301 & no var & SB? \\
MHF81906a &05 30 04.67& -67 18 39.60 & 14.90 &      &      & +415 & +323 & no var & SB2 \\
MHF81906b &            &              &       &      &      & +186 & +384 & &  \\
MHF83937a &05 29 33.98&-67 18 06.80& 14.87 & +305 & +265 & +305 & +296 & no var & SB2 \\
MHF83937b &             &              &      & +305 & +350 & +305& +296 &  &  \\
MHF87970a &05 29 08.34& -67 17 18.30 & 15.10 & +263 & +403 & +340 & +257 &7.117 & SB2 \\
MHF87970b &            &              &       & +496 & +121 & +280 & +512?& &  \\
MHF91603 &05 28 03.40 & -67 16 33.20 & 15.19 & +318 & +291 & +321 & +300 & no var & SB? \\
MHF94292a &05 28 24.68& -67 16 04.10 & 15.26 & +428 & +367 & +199 & +412 & no var & SB2 \\
MHF94292b &            &              &       & +164 & +220 & +390 & +172 & &  \\
MHF98013 &05 30 32.65& -67 15 25.70 & 14.82 & +267 & +330 & +265 & +322 & no data &  \\
MHF102053&05 29 56.24&-67 14 32.95& 16.26 & +349 & +295 & &      & no data &  \\
MHF103207&05 30 13.56& -67 14 27.10 & 14.94 &      &      & +205 & +444 & no data &   \\
MHF109251&05 28 31.95&-67 13 11.90& 15.95 &      &     & +283& +249 & no var & SB? \\
MHF110467&05 27 25.31& -67 12 52.50 & 15.70 &     &      & +266 & +323 & no var & \\
MHF111340a&05 27 14.43& -67 12 40.90& 16.16 & +453 & +167 & +492 & +263 &1.074 & SB2 \\
MHF111340b&            &             &       & +142 & +453 & +70  & +361 & &  \\
MHF112849a&05 29 54.86& -67 12 41.00 & 14.81 & +358 & +299 & +227 & +262 & no data & SB2 \\
MHF112849b&            &             &       & +115 & +270 & +461 & +315 &   &  \\
MHF113048 &05 26 57.53& -67 12 18.95 & 16.38 & +283 & +251 & &    & no var & SB? \\
MHF127573a &05 27 19.79& -67 09 34.60& 16.46 & +378 & +260 & +371 & +217 &2.932 & SB2 \\
MHF127573b &            &             &       & +100 & +260 & +163 & $\sim +300$ & &  \\
MHF128963 & 05 30 38.29 & -67 09 32.70&15.95 &      &      & +315 & +274 & no var & SB? \\
MHF133975 &05 29 50.61& -67 08 30.00& 16.27 & +241 & +352 & +233 & +344 & no data & \\
MHF136274a &05 28 13.38& -67 07 58.50& 16.28 &  +365 & +276 & +292& +287 & no var &SB2 \\
MHF136274b &            &             &       & +163 & +276 & +292 & +287 & &  \\
MHF141891 &05 28 00.77& -67 06 55.50& 16.29 & +230 & +288 & +264 & +231 &2.975 &  \\
MHF149652 &05 28 58.67& -67 05 29.20& 16.50 & +338 & +276 & +261 & +295 &1.458 &  \\
\hline
	\end{tabular}
	\label{table2}
\end{table*}

The radial velocity and the strength of nebular lines (H$\alpha$,
[\ion{N}{ii}] and [\ion{S}{ii}]) have been determined, as well as the radial
velocity of strong photospheric lines (H$\alpha$, H$\gamma$, H$\delta$,
\ion{He}{i} and \ion{Mg}{ii}). The accuracy is $\pm$10~km~s$^{-1}$. The
statistical distribution of stellar and nebular RVs is given in
Fig.~\ref{figure2}. The distribution is Gaussian for stellar RVs with a
maximum around +300 km~s$^{-1}$ while it is bi-modal for nebular RVs with
peaks at +305 and +335 km~s$^{-1}$. After comparing stellar and nebular RVs for each star, 
no clear link is found between the stellar object and the nebular 
emission in its spectrum.

Thus, there is apparently no physical link between stars and the structures 
giving rise to the nebular lines.

In Fig.~\ref{figure3} we plot the position of stars, with and without nebular
lines, and of sky fibres in the field. Different symbols are used to
distinguish stars  in which the nebular emission has low or high RVs, 
as well as stars with strong nebular lines.
From Fig.~\ref{figure3}, it can be seen that high and low nebular RVs are not
randomly distributed over the field but seem to be organized in structures
that look like filaments. The higher velocities centered around  +335
km~s$^{-1}$ are observed in the eastern and northwestern parts of the field,
while the lower velocities, centered around +305 km~s$^{-1}$, are observed in
the southwestern part. Higher intensity is generally found in nebulosities
with the lowest H$\alpha$ radial velocities. However, stars with no nebular
H$\alpha$ line, also indicated in Fig.~\ref{figure3}, are found in isolated
regions of the observed field, reflecting the patchy nature of nebulosities in
this part of the LMC. 

The mean value of the [\ion{S}{ii}] 6717/6731 ratio is 1.4, typical of LMC
bubbles (Skelton et al. 1999). The [\ion{N}{ii}]/H$\alpha$ ratio is  lower than
0.1; the S[II](6717 or 6731 \AA)/H$\alpha$ ratio ranges between 0.1 and 0.3.
These values are close to those found for the \ion{H}{ii} regions in
the LMC.

The patchy distribution with a variable density, seen in the analysis of
the nebular lines H$\alpha$, [\ion{N}{ii}] and [\ion{S}{ii}], agrees 
fairly well with was the results of \ion{H}{i}
distribution survey (Staveley-Smith et al. 2003). Indeed, the \ion{H}{i}
distribution has revealed that the body of the LMC is punctuated by large holes
and has a generally mottled appearance. One of the main \ion{H}{i} gaps in the
LMC corresponds to LMC4. The southern inner limit of the LMC4 supergiant shell,
taken from Kim et al. (1999), crosses the field we observed with the
VLT-GIRAFFE multi-object spectrograph (see Fig.~\ref{figure3}).
The LMC4 \ion{H}{i} supergiant shell is centered at $\alpha$(2000)=05h 31min
33s and $\delta$(2000) = -66$^{\circ}$  40$\arcmin$  28$\arcsec$, its shell
radius is r(sh)~=~38.7$\times$37.7$\arcmin$ and its systemic heliocentric
velocity is RV~=~+306~km~s$^{-1}$ (Kim et al. 1999); note that the shell radius
previously determined by Meaburn (1980) is larger by 12\arcmin.

Our study of nebular lines shows that:
\begin{itemize}
\item  Higher-velocity nebulosities (RVnebH$\alpha$=+335 km~s$^{-1}$ on
average) appear to coincide with the southern inner \ion{H}{i} gas in LMC4.
\item Lower-velocity nebulosities (RVnebH$\alpha$=+305 km~s$^{-1}$ in average)
are preferentially detected at the rim of LMC4 and have the same RV as the
systemic RV of the bubble.
\item The separation between lower and higher velocities roughly corresponds
to the inner limit of the LMC4 \ion{H}{i} supergiant shell.
\item The strong intensity of nebular lines observed in the southern
(S-W) part of the field is easily explained by the vicinity of the
\ion{H}{ii}
region LHA 120-N 51A.
\end{itemize}
\subsection{Binaries}
Discrepancies between stellar radial velocities of blue and red lines
(interval of 4 days in November 2003 and 2 days in April 2004) greater than 
2$\sigma$~=~18~km~s$^{-1}$ allow us to detect 18 binaries and to suspect 5 
others of binarity (see Table~\ref{table2}). 
The binary nature of the latter objects  may be questionable because the 
discrepancies observed in the
radial velocities are between 2 and 3$\sigma$ and further observations are
therefore needed to confirm their binary nature. Among the 23 
binaries, 8 objects are SB2. 

Despite the small number of spectra obtained for each object, it is
possible to roughly estimate the systemic velocity and the mass ratio m1/m2
for the SB2 systems using a graphical method (Wilson 1941) based on the linear
relation between the velocity of each component of the system. Results are
given in Table~\ref{table3}. 

In addition, we searched for light curves of the new SB systems in the MACHO
database. All the sources in the MACHO database have been identified thanks to 
their GIRAFFE coordinates and fall inside a radius smaller than 1\arcsec. Among
the 18 identified sources, 5 stars are found to be eclipsing binaries:
MHF87970 (SB2), MHF111340 (SB2), MHF127573 (SB2), MHF141891 (SB1) and MHF149652
(SB1); two of them seem to show a total eclipse (MHF127573 and MHF149652). For
these eclipsing binaries the orbital period is determined using 3 different
methods: the CLEAN algorithm (Roberts et al. 1987), the Least Squares
method and the PDM method (Stellingwerf 1978). The periods are given
in Table~\ref{table2}.

Illustrations of the most characteristic systems, associated with their MACHO
light curve when available, are given in Figs~\ref{83937} to \ref{149652}. The
light curve magnitudes are expressed in instrumental units.

\begin{itemize}
\item{MHF83937, MACHO 60.8070.1685} (see Fig.~\ref{83937}): In the blue 
range, spectral lines
show a single component only in November 2003 and April 2004. 
Splitting is observed in the red range in November 2003, but in April 2004 the
red lines become deeper and again single, which indicates an eclipse phase. The
mass ratio is about 1. This star has been observed by MACHO but no periodicity
could be found.
\\
\begin{figure}[!]
	\centering

\resizebox{\hsize}{!}{\includegraphics[angle=270]{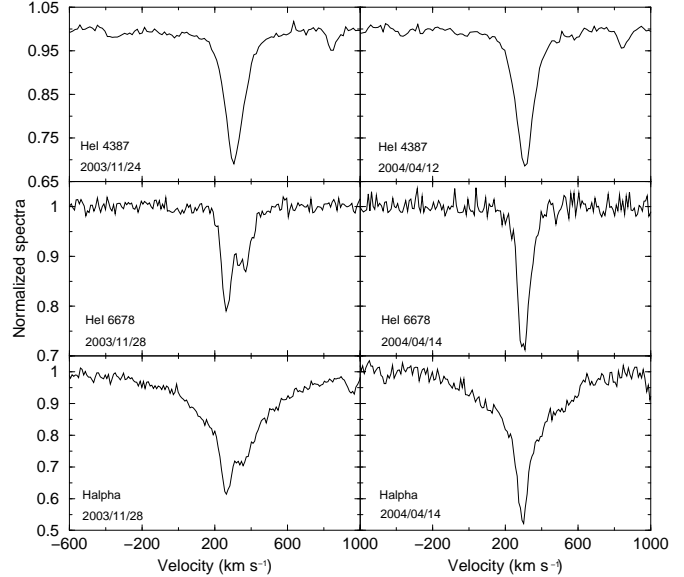}}
	\caption{Spectroscopic VLT-GIRAFFE observations of the binary
star MHF83937}
	\label{83937}
\end{figure}
\item{MHF87970, MACHO 60.7950.16} (see Fig.~\ref{87970}): An
interchange of position between the primary and secondary line
profiles was observed at both epochs of observations (2003 and 2004). The
estimated mass ratio is about 2.8. The deeper and narrower component  is
attributed to the secondary star. Note that the \ion{Mg}{ii} 4481 line
always shows a single component with the same RV as the secondary
star. This secondary is probably a giant A star. From MACHO data it was
possible to determine the orbital period P$_{\rm orb}$ = 7.117 d. The light
curve indicates that the system is an eclipsing one with 2 roughly identical
minima, which means that the 2 components of the system have 
approximately the same luminosity.\\
\begin{figure}[!]
	\centering

\resizebox{\hsize}{!}{\includegraphics[angle=270]{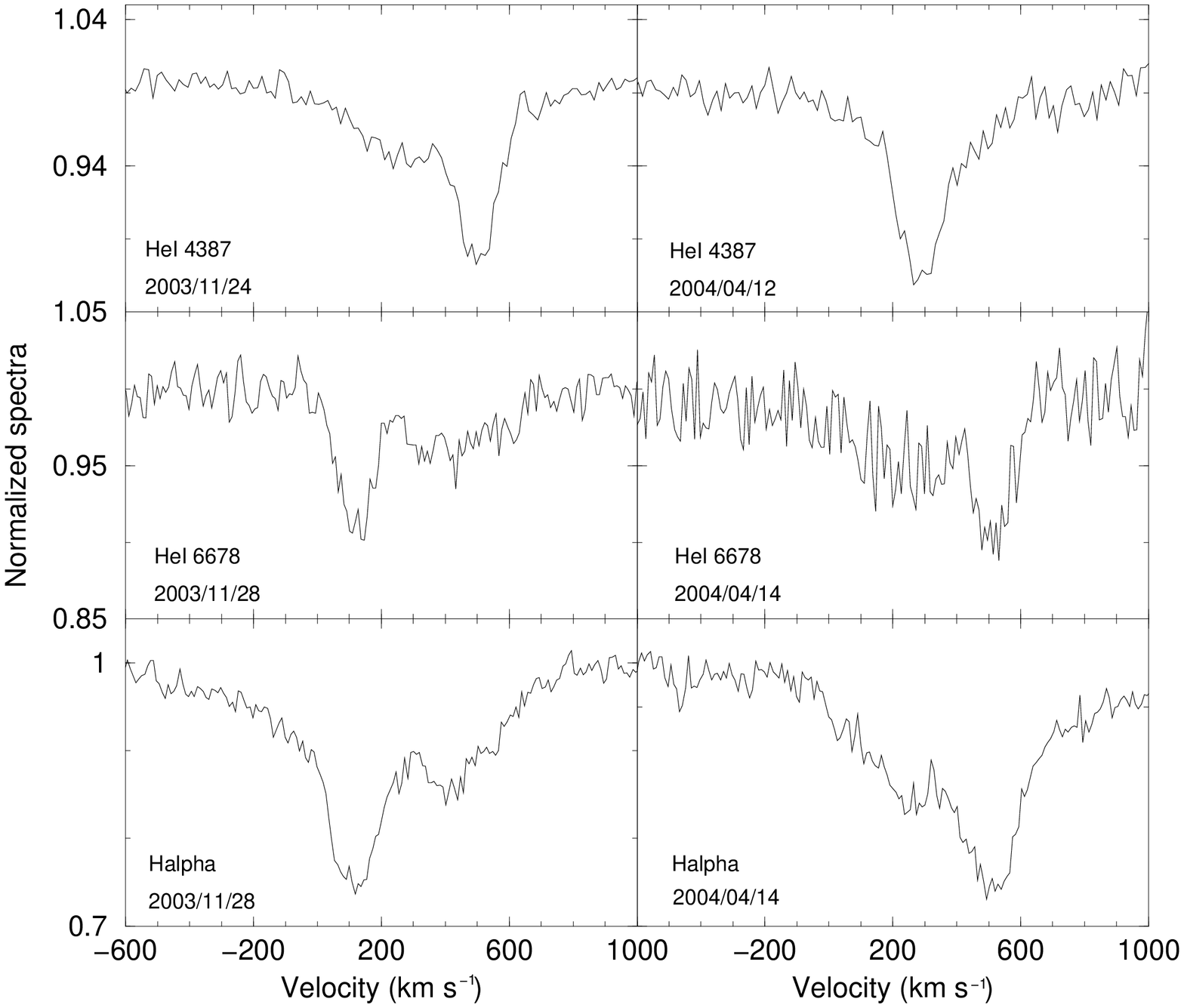}}

\resizebox{\hsize}{!}{\includegraphics[angle=270]{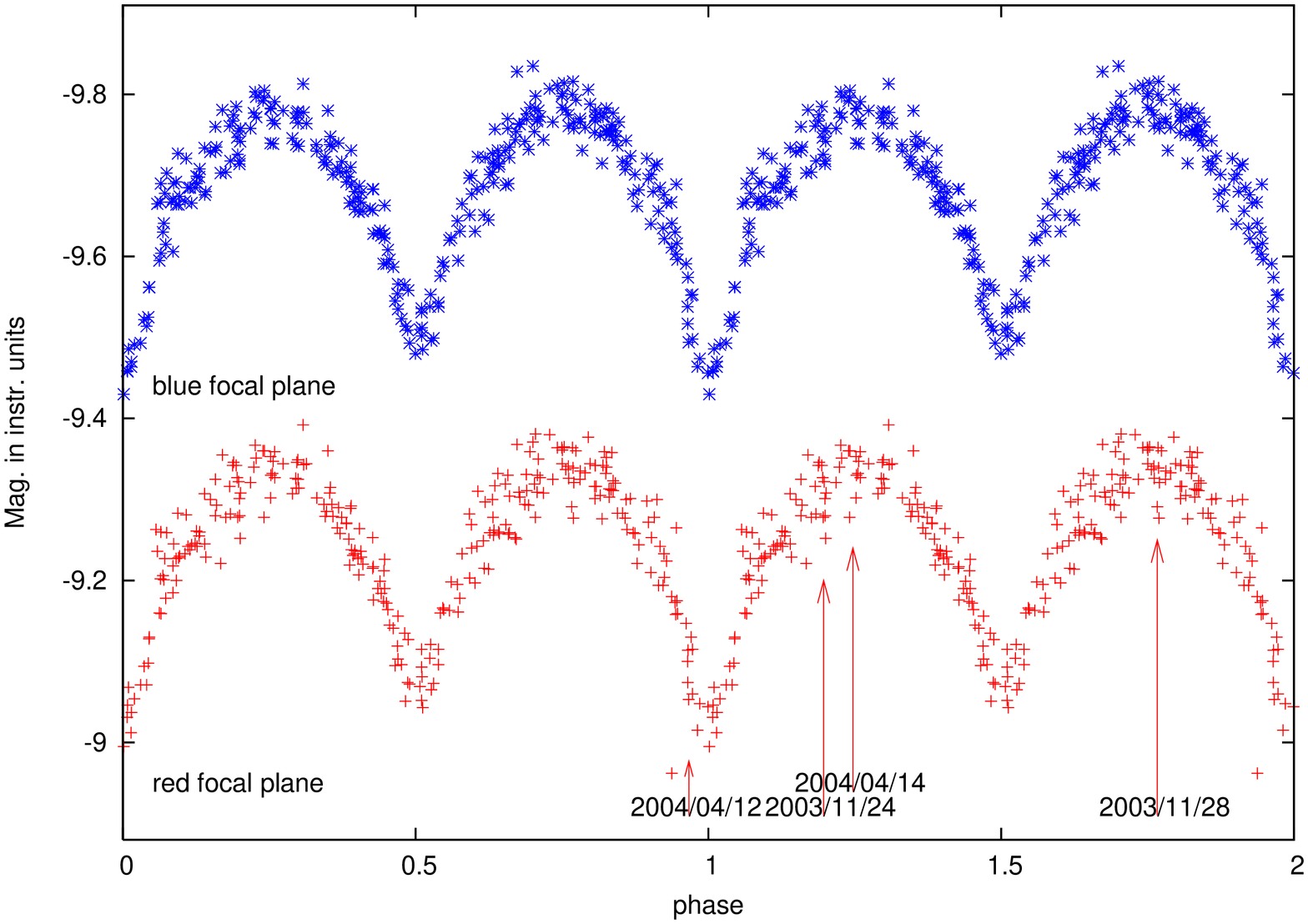}}
	\caption{Observations of the binary star MHF87970. Top: spectroscopic
	VLT-GIRAFFE data; bottom: MACHO data folded in phase with
	P$_{\rm orb}$ = 7.117 d.}
	\label{87970}
\end{figure}
\item{MHF94292, MACHO 60.7629.23} (see Fig.~\ref{94292}): Hydrogen 
and helium lines are double in
each spectrum and line profile interchanges of position are observed 
over 2 days in 2004.
This implies that the orbital period is a few days but no periodicity 
can be determined from MACHO data. The mass ratio is about 1.
\\
\begin{figure}[!]
	\centering

\resizebox{\hsize}{!}{\includegraphics[angle=270]{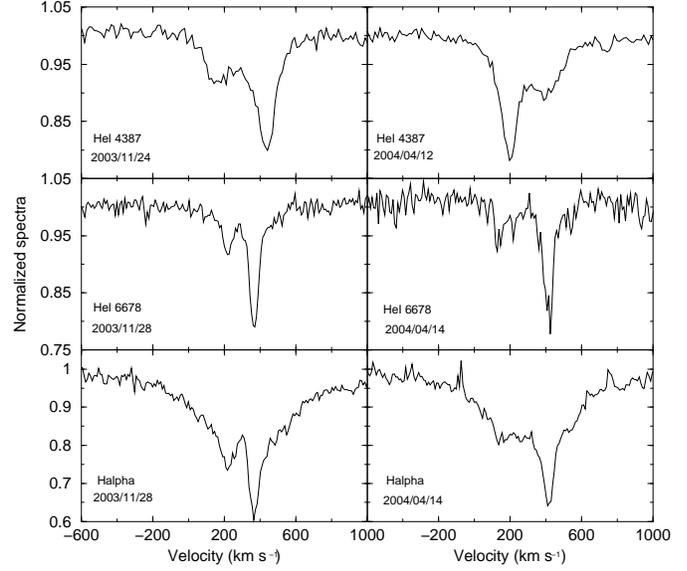}}
	\caption{Spectroscopic VLT-GIRAFFE observations of the binary
star MHF94292.}
	\label{94292}
\end{figure}
\begin{table}[htb]
	\centering
	\caption{Systemic radial velocity and mass ratio of the binary SB2 systems discovered
in the region surrounding of NGC2004. Accuracy is about
$\pm$20 km~s$^{-1}$ on RVsyst and $\pm0.4$ for mass ratio.}
	\begin{tabular}{lll}
\hline
\hline
Binary & RVsyst & m1/m2  \\
     & km~s$^{-1}$   &       \\
\hline
MHF81906  & +311 & 1.3 \\
MHF83937 & +307 & $\sim 1$ \\
MHF87970  & +338 & 2.8 \\
MHF94292  & +294 & 1.0 \\
MHF111340 & +304 & 1.2 \\
MHF112849 & +289 & 2.3 \\
MHF127573 &  +260  &  $\sim 1$ \\
MHF136274 & +290  & $\sim 1.7$ \\
\hline
	\end{tabular}
	\label{table3}
\end{table}
\item{MHF111340, MACHO 60.7566.36} (see Fig.~\ref{111340}): The blue spectrum
in 2003 and 2004 shows 2 widely separated components. In 2003 the red spectrum,
obtained 4 days after the blue one, displays an interchange in the
component positions. In 2004 the red spectrum, taken 2 days after the blue
one, probably shows an interchange of position as well, but it is at
the limit of detectability due to a poor S/N ratio and to the presence of a
strong nebular line at H$\alpha$. Estimated mass ratio is about 1.2.
A short orbital period was found in the MACHO data, P$_{\rm orb}$ =
1.074 d.
\\
%
\begin{figure}[!]
	\centering
		 \resizebox{\hsize}{!}{\includegraphics{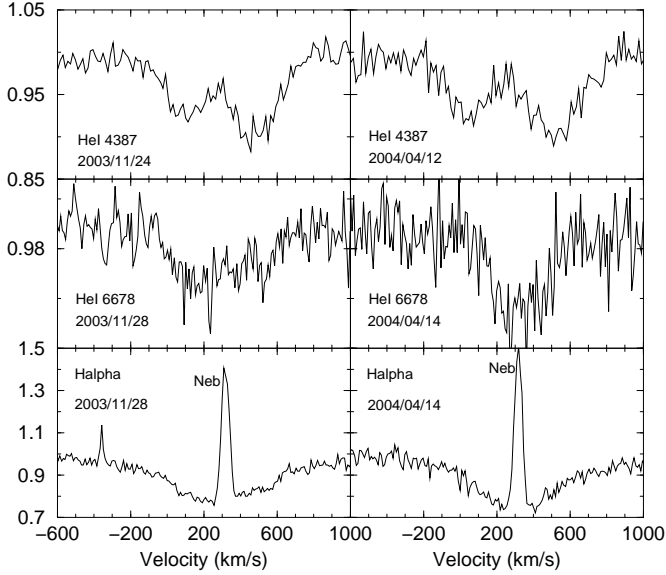}}

\resizebox{\hsize}{!}{\includegraphics[angle=270]{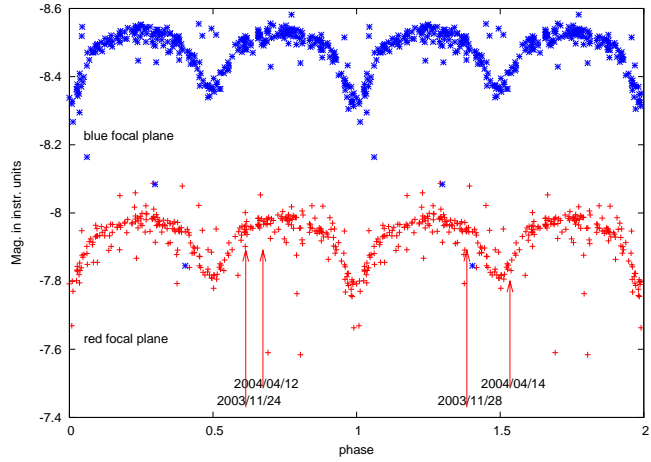}}
	\caption{Observations of the binary star MHF111340. Top: spectroscopic
	VLT-GIRAFFE data; bottom: MACHO data folded in phase with
	P$_{\rm orb}$ = 1.074 d.}
	\label{111340}
\end{figure}
\item{MHF127573, MACHO 60.7710.39} (see Fig.~\ref{127573}): Blue \ion{He}{i} 
lines are double in
November 2003 and April 2004 and red ones are very noisy but shifted
compared to the blue ones observed 4 and 2 days before, respectively.
H$\alpha$ is highly disturbed by a nebular component but is fainter and
broader in April 2004 than in November 2003. This suggests a splitting of
the line  due to the secondary at about +300 km~s$^{-1}$. The orbital period
determined from the MACHO data is 2.932 d. Flat-bottomed light curve 
indicates a total
eclipse with 2 slightly unequal minima.
The mass ratio is about 1.
\\
\begin{figure}[!]
	\centering
		 \resizebox{\hsize}{!}{\includegraphics{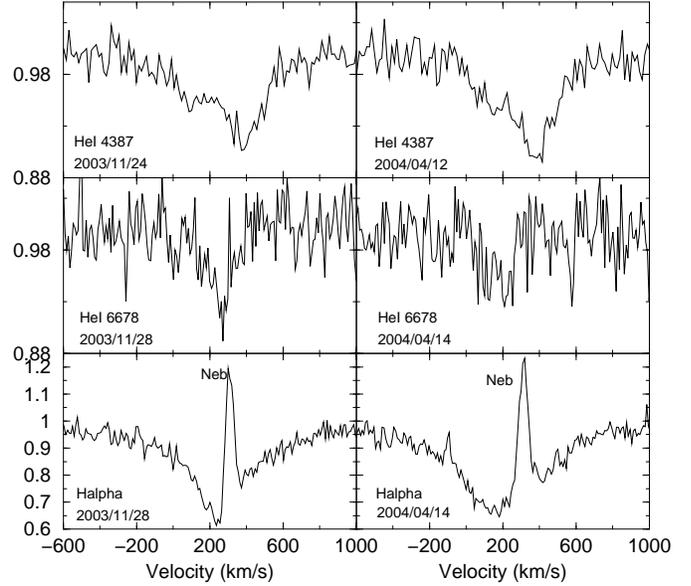}}

\resizebox{\hsize}{!}{\includegraphics[angle=270]{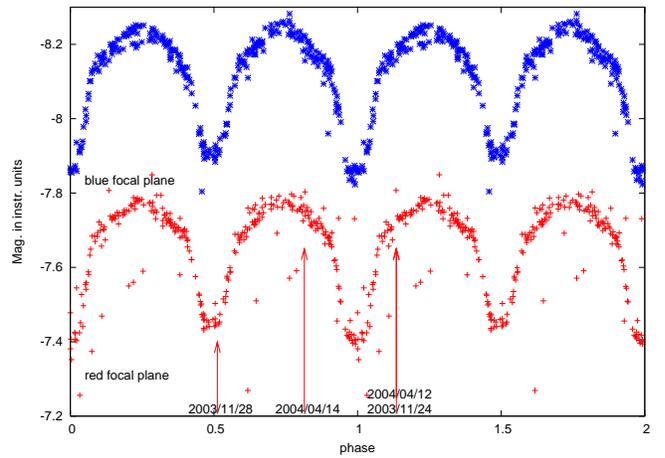}}
	\caption{Observations of the binary B star MHF127573. Top:
spectroscopic VLT-GIRAFFE data; bottom: MACHO data folded in phase
with P$_{\rm orb}$ = 2.932 d.}
	\label{127573}
\end{figure}
\item{MHF141891, MACHO 60.7710.36} (see Fig.~\ref{141891}): This B-type star 
is a SB1 system. A
faint nebular H$\alpha$ line disturbs the photospheric line profile, but it
seems that the profile obtained in 2004 is larger, in equivalent width, 
than the profile obtained in
2003. This could indicate that the secondary spectrum is at the limit of
detection. The MACHO data allow the determination of an orbital period of
2.975 d, the light curve being typical of an eclipsing binary with 2 unequal
minima. The eclipse is partial.
\\
\begin{figure}[!]
	\centering
\resizebox{\hsize}{!}{\includegraphics[angle=270]{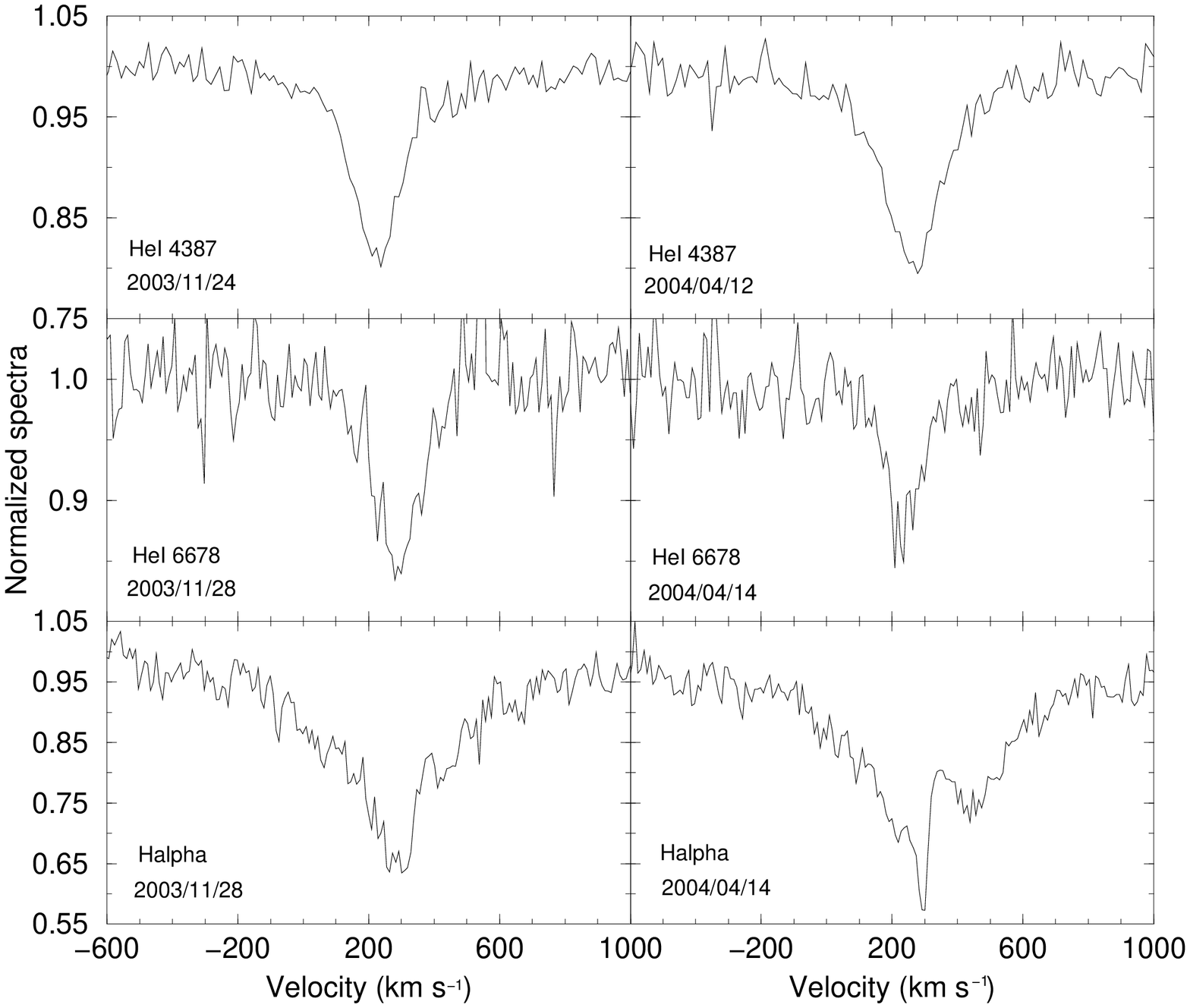}}

\resizebox{\hsize}{!}{\includegraphics[angle=270]{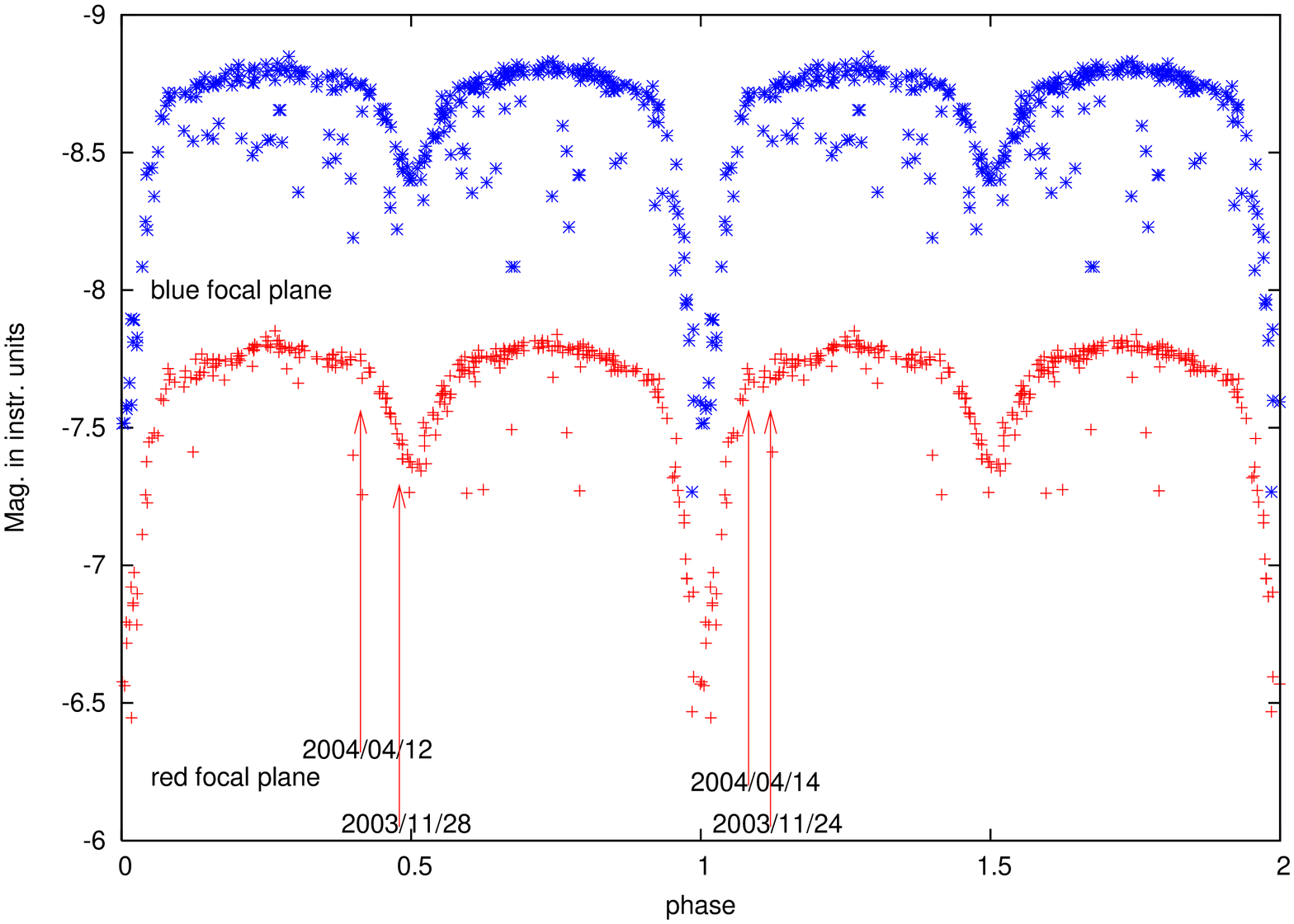}}
	\caption{Observations of the SB1 B star MHF141891. Top: spectroscopic
	VLT-GIRAFFE data; bottom: MACHO data folded in phase with
	P$_{\rm orb}$ = 2.975 d.}
	\label{141891}
\end{figure}
\item{MHF149652, MACHO 60.7953.36} (see Fig.~\ref{149652}): This binary system 
always shows a
single spectrum on our spectroscopic observations. However, the light curve
obtained from the MACHO data folded in phase with the orbital period P$_{\rm
orb}$ = 1.458 d is typical of a total eclipsing binary with 2 unequal minima.

\begin{figure}[!]
	\centering
\resizebox{\hsize}{!}{\includegraphics[angle=270]{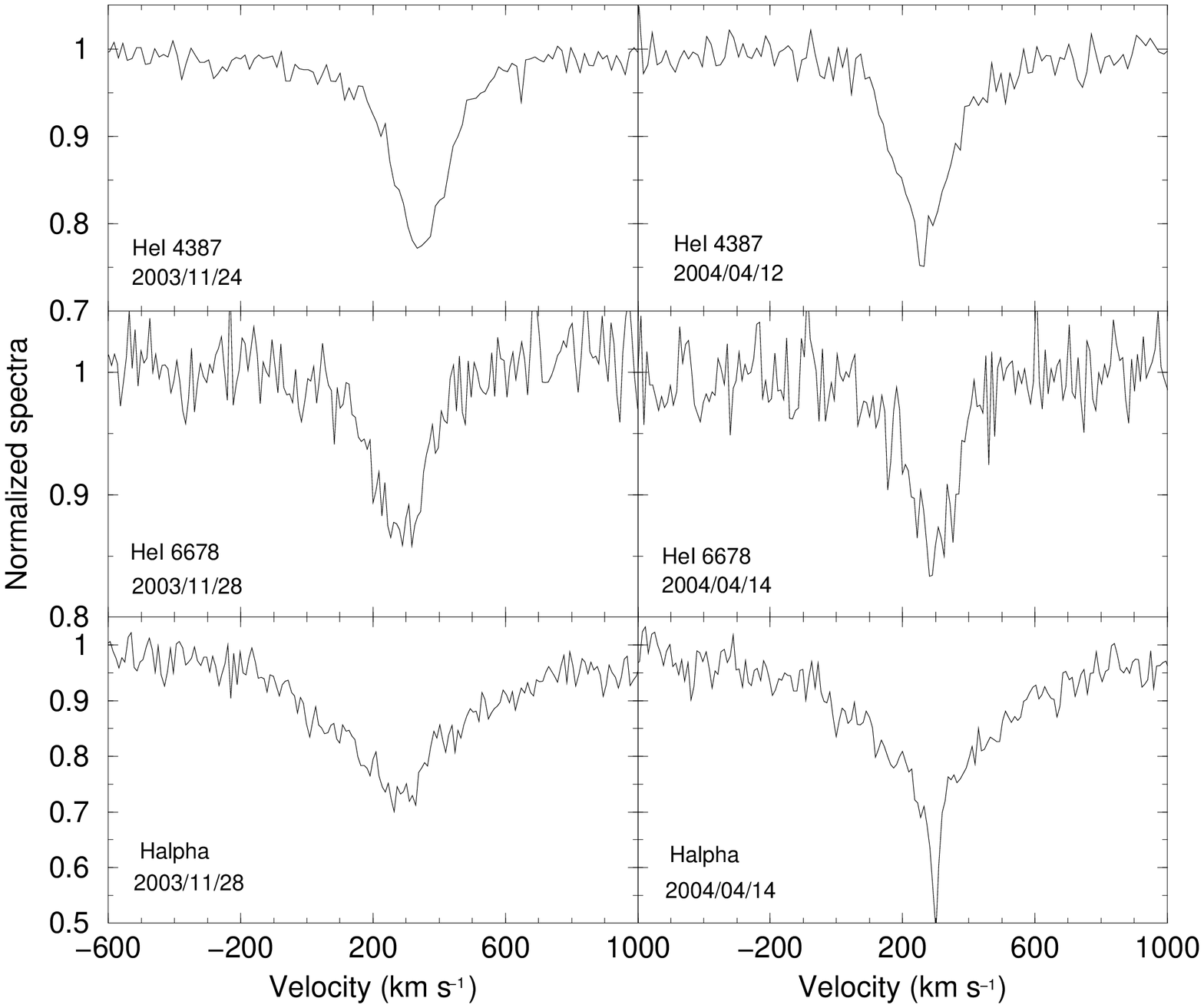}}

\resizebox{\hsize}{!}{\includegraphics[angle=270]{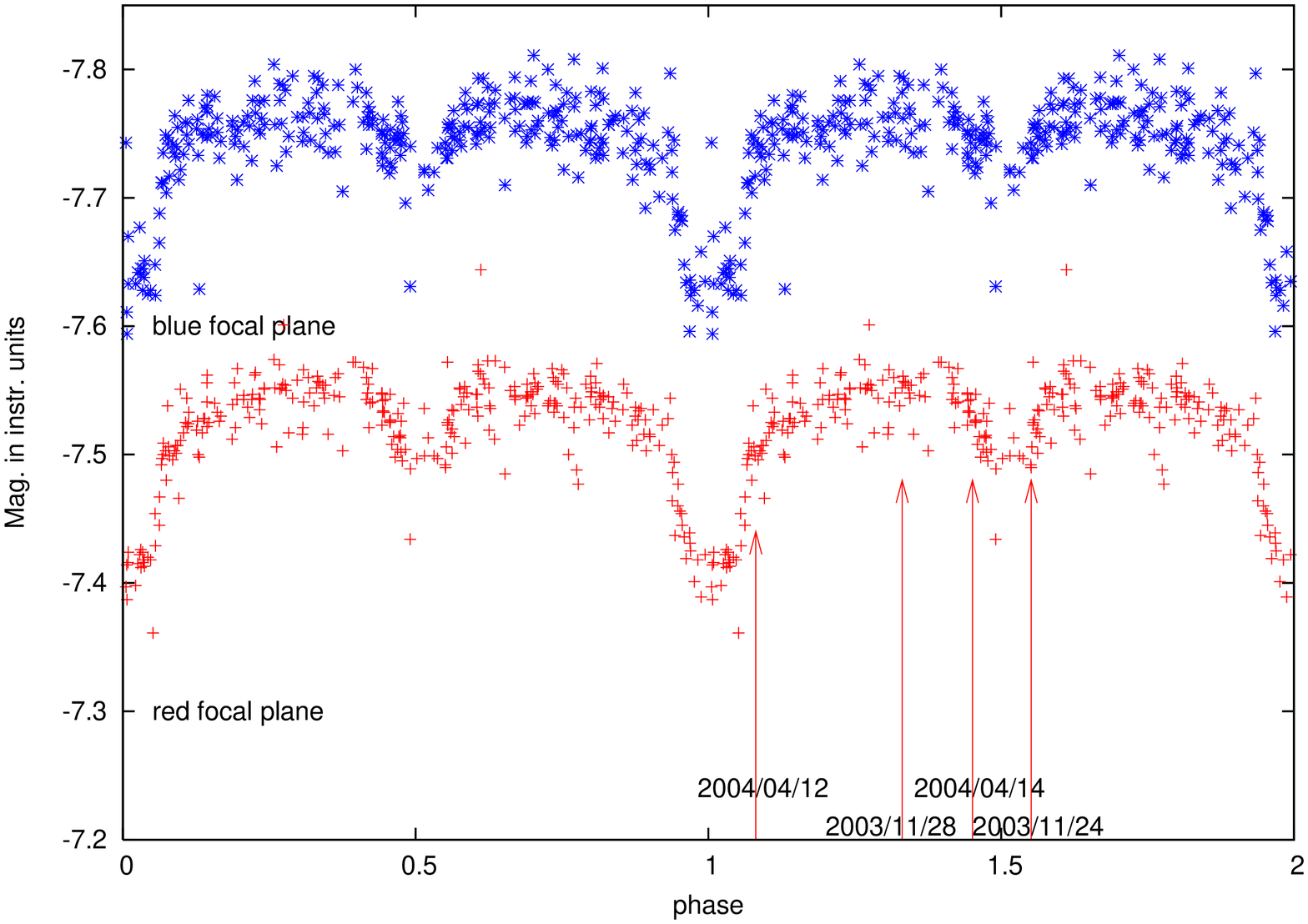}}
	\caption{Observations of the SB1 B star MHF149652. Top: spectroscopic
	VLT-GIRAFFE data; bottom: MACHO data folded in phase with
	P$_{\rm orb}$ = 1.458 d.}
	\label{149652}
\end{figure}

\end{itemize}
The percentage of binaries among our sample is $\sim$12\%. This is probably a
lower limit, since only systems with relatively short periods can be
detected with our observations. In the Galaxy the rate of spectroscopic
binaries is $\sim$20\% for early-type stars.
\section{Conclusion}
For the first time, medium resolution spectroscopic observations of a large
sample of B and Be stars in the LMC-NGC2004 region are
presented. 25 new Be stars have been identified in the field 
as well as small clusters and OB concentrations.

The high capability of the VLT-GIRAFFE multi-object spectrograph
allowed us to identify nebular lines that pollute 75\% of the 
star  spectra observed.  Careful study enabled us to disentangle
these lines from the CS contribution in Be stars. We were therefore 
able to 
separate Be stars with genuine single-peak from double-peak H$\alpha$
emission lines. Characteristics of the H$\alpha$ emission line have been
investigated through  spectral parameters (FWHM, EW$\alpha$, peak separation).
No noticeable difference was found between the LMC and galactic Be
stars.

The proportion of Be stars compared to B-type stars was found to be of
the same order in the   Large Magellanic Cloud and in the Galaxy fields. Thus
metallicity does not seem to strongly influence the apparition of
the Be phenomenon.

A by-product of the observations of B-type stars in the LMC region is
the study of nebular H$\alpha$, [\ion{S}{ii}] and [\ion{N}{ii}] lines. We
found that the nebulosities in which these lines are formed are not
related to the B and Be stars of the sample. The RV distribution of
these nebular lines is bi-modal and peaks at +305 and +335 km~s$^{-1}$, while
it is Gaussian for the stellar lines. The nebular structures located in the
north-eastern part of the field observed with VLT-GIRAFFE have higher RV (+335
km~s$^{-1}$) and lower density than those located in the southern-western
part (+305 km~s$^{-1}$) next to the \ion{H}{ii} region LHA 120-N51A.
The patchy nature of the nebulosities in the LMC, already detected on 
a large
scale by \ion{H}{i} surveys, is widely confirmed on a small scale by this study. 

Finally, we identified 23 new binary systems, of which 5 have still to be confirmed.
8 systems are SB2 and their systemic RV and mass ratio have been derived.
Using the MACHO database, 5 systems have been found to be eclipsing
binaries, which allows us to accurately determine their orbital
period. The systems MHF111340 and MHF149642 have very short orbital periods: 
1.074 and 1.458 d
respectively. Of the 8 SB2 we detected from VLT-FLAMES 
spectra
and the 5 systems we identified with MACHO light curves, three are in common 
(MHF87970, MHF111340, MHF136274).

A study of the fundamental parameters of the B and Be stars of this sample will
be presented in a forthcoming paper (Martayan et al., in preparation).

\begin{acknowledgements}
We would like to thank Dr H. Flores for the success and good quality 
of the observing run in November 2003.
We thank Drs M.R. Cioni and J. Smoker for their help 
during the observing run in April 2004.
We thank M. Mekkas for his technical support. 
We also thank the referee Dr T. Rivinius for his constructive remarks.
This paper utilizes public domain data originally obtained by MACHO Project,
whose work was performed under the joint auspices of the U.S. Department of
Energy, National Security Administration by the University of California,
Lawrence Livermore National Laboratory under contract No. W-7405-Eng-48, the
National Science Foundation through the Center for Particle Astrophysics of
the University of California under cooperative agreement AST-8809616, the
Mount Stromlo and Siding Spring Observatory, part of the Australian National
University.

This research has made use of the Simbad database and Vizier database maintained 
at CDS, Strasbourg, France.

\end{acknowledgements}

\end{document}